\documentclass{iaus238}
\usepackage{graphicx}

\title[Strong gravity effects] 
{Strong gravity effects: X-ray spectra, variability and polarimetry}

\author[A.~C. Fabian]   
{Andrew~C. Fabian}

\affiliation{Institute of Astronomy, Madingley Road,
Cambridge CB3 0HA, UK \\[\affilskip]
email: acf@ast.cam.ac.uk}

\pubyear{2006}
\volume{238}  
\pagerange{001--999}
\date{??? and in revised form ???}
\jname{Black Holes: from Stars to Galaxies -- across the Range of Masses}
\editors{V. Karas \& G. Matt, eds.}

\begin{document}

\maketitle

\begin{abstract}
Accreting black holes often show iron line emission in their
X-ray spectra. When this line emission is very broad or variable then
it is likely to originate from close to the black hole. The theory and
observations of such broad and variable iron lines are briefly
reviewed here. In order for a clear broad line to be found, one or more
of the following have to occur: high iron abundance, dense disk
surface and minimal complex absorption. 

Several excellent examples are found from observations of Seyfert 
galaxies and Galactic Black Holes. In some cases there is strong 
evidence that the black hole is rapidly spinning. 
Further examples are expected as more long observations are 
made with XMM-Newton, Chandra and Suzaku. The X-ray spectra show
evidence for the strong gravitational redshifts and light bending
expected around black holes.

\end{abstract}

\firstsection 

\maketitle

\section{Introduction}

Much of the radiation from luminous accreting black holes is released
within the innermost 10-20 gravitational radii (i.e. 
$10$--$20 r_{\rm{}g}\equiv 10$--$20 GM/c^2$). In such an energetic environment, iron 
is a major source of line emission, with strong emission lines in the
6.4--6.9~keV band. Observations of such line emission then provides us
with a diagnostic of the accretion flow and the behaviour of matter
and radiation in the strong gravity regime very close to the black
hole (Fabian \etal\ 2000; Reynolds \& Nowak 2003; Fabian \& Miniutti
2005).

The rapid X-ray variability found in many Seyfert galaxies is strong
evidence for the emission orginating at small radii. The high
frequency break in their power spectra, for example, corresponds to
orbital periods at $\sim20 r_{\rm g}$ and variability is seen at still
higher frequencies (Uttley \& McHardy 2004; Vaughan \etal\ 2004). Key
evidence that the very innermost radii are involved comes from
Soltan's (1982) argument relating the energy density in radiation from
active galactic nuclei (AGN) to the local mean mass density in massive
black holes, which are presumed to have grown by accretion which
liberated that radiation. The agreement found between these quantities
requires that the radiative efficiency of accretion be 10 per cent or
more (Yu \& Tremaine 2002; Marconi \etal\ 2005). This exceeds the 6 per
cent for accretion onto a non-spinning Schwarzschild black hole and
inevitably implies that most massive black holes are rapidly spinning
with accretion flows extending down to a just few $r_{\rm g}$.
Moreover, this is where most of the radiation in such accretion flows
originates.

\begin{figure}[tbh]
\includegraphics[width=0.65\textwidth]{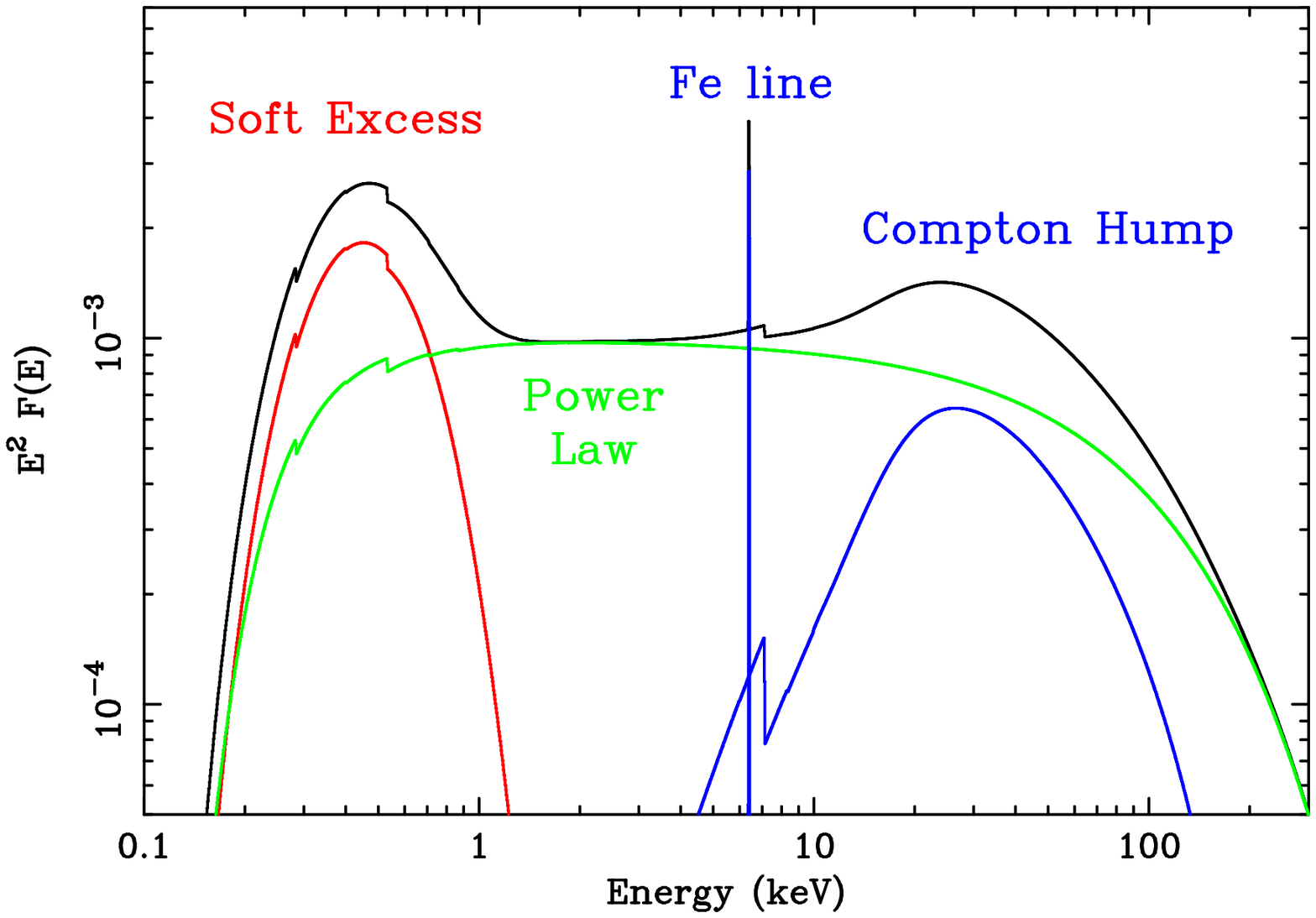}
\hspace{0.1\textwidth}
\includegraphics[width=0.4\textwidth]{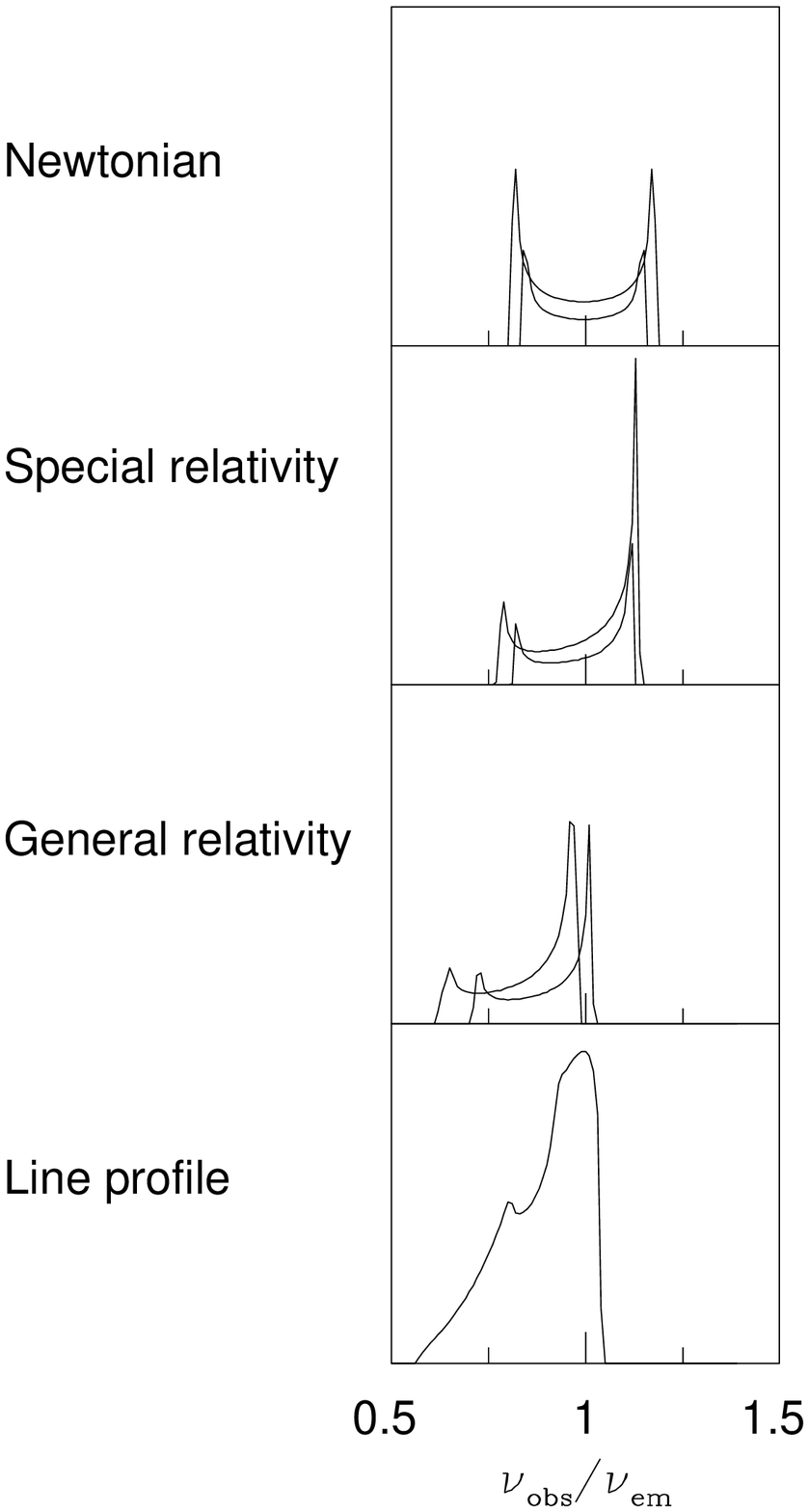}
\caption{Left: The main components of the X--ray spectra of
  unobscured accreting BH are shown: soft X--ray emission from the
  accretion disc (red); power law from Comptonization of the soft
  X--rays in a corona above the disc (green); reflection continuum and
  narrow Fe line due to reflection of the hard X--ray emission from
  dense gas (blue). Right: The profile of an intrinsically narrow
  emission line is modified by the interplay of Doppler/gravitational
  energy shifts, relativistic beaming, and gravitational light bending
  occurring in the accretion disc (from Fabian \etal\ 2000). The upper
  panel shows the symmetric double--peaked profile from two annuli on
  a non--relativistic Newtonian disc. In the second panel, the effects
  of transverse Doppler shifts (making the profiles extend to lower
  energies) and of relativistic beaming (enhancing the blue peak with
  respect to the red) are included. In the third panel, gravitational
  redshift is turned on, shifting the overall profile to the red side
  and reducing the blue peak strength. The disc inclination fixes the
  maximum energy at which the line can still be seen, mainly because
  of the angular dependence of relativistic beaming and of
  gravitational light bending effects.  All these effects combined
  give rise to a broad, skewed line profile which is shown in the last
  panel, after integrating over the contributions from all the
  different annuli on the accretion disc. Detailed computations are
  given by Fabian \etal\ (1989), Laor (1991), Dov\v{c}iak \etal\ (2004) and
  Beckwith \& Done (2004). }
\end{figure}

The X-ray spectra of AGN are characterized by several components: a
hard power-law which may turnover at a few hundred keV, a soft excess
and a reflection component (Fig.~1). This last component is produced
from surrounding material by irradiation by the power-law. It consists
of backscattered X-rays, fluorescence and other line photons,
bremsstrahlung and other continua from the irradiated surfaces.
Examples of reflection spectra from photoionized slabs are shown in
Fig.~2. At moderate ionization parameters ($\xi=F/n\sim
100$~erg~cm~s$^{-1}$, where $F$ is the ionizing flux and $n$ the
density of the surface) the main components of the reflection spectrum
are the Compton hump peaking at $\sim 30$~keV, the iron line at
6.4--6.9~keV (depending on ionization state) and a collection of lines and
reradiated continuum below 1~keV. When such a spectrum is produced
from the innermost parts of an accretion disk around a spinning black
hole, the outside observer sees it smeared and redshifted (Fig.~2)
due to doppler and gravitational redshifts.

Another, potentially powerful, diagnostic of the strong gravity regime
is the study of timing and Quasi-Periodic Oscillations (QPOs).
Despite the richness of the data (e.g. Strohmayer 2001), there is no
consensus on how to interpret them and they will not be discussed
further here.

\begin{figure}[tbh]
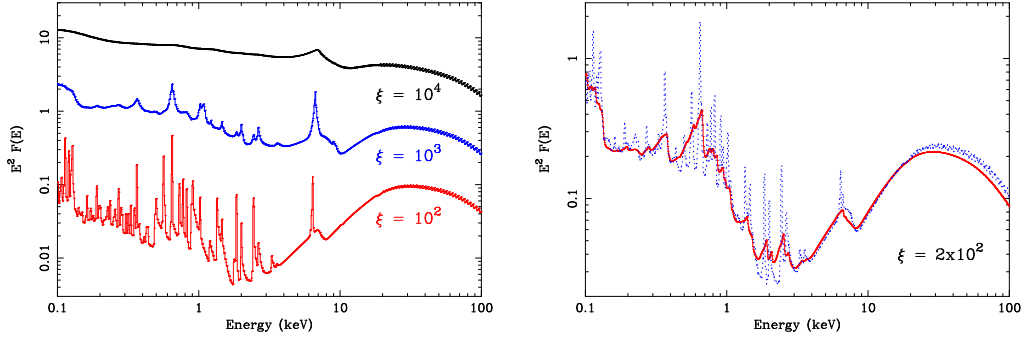

\includegraphics[height=0.48\textwidth,angle=-90]{rossreflion.ps}
\hfill
\includegraphics[height=0.48\textwidth,angle=-90]{refblur.ps} 
\caption{Left: Computed X-ray reflection
    spectra as a function of the ionization parameter $\xi$ (from the
    code by Ross \& Fabian 2005).  The illuminating continuum has a
    photon index of $\Gamma=2$ and the reflector is assumed to have
    cosmic (solar) abundances. Right: Relativistic effects
    on the observed X-ray reflection spectrum (solid line). 
    We assume that the intrinsic
    rest-frame spectrum (dotted) is emitted in an accretion disc
    and suffers all the relativistic effects shown in Fig.~1. } 
\end{figure}

\section{Observations}
\begin{figure}[tbh]
\includegraphics[height=0.51\textwidth,angle=-90]{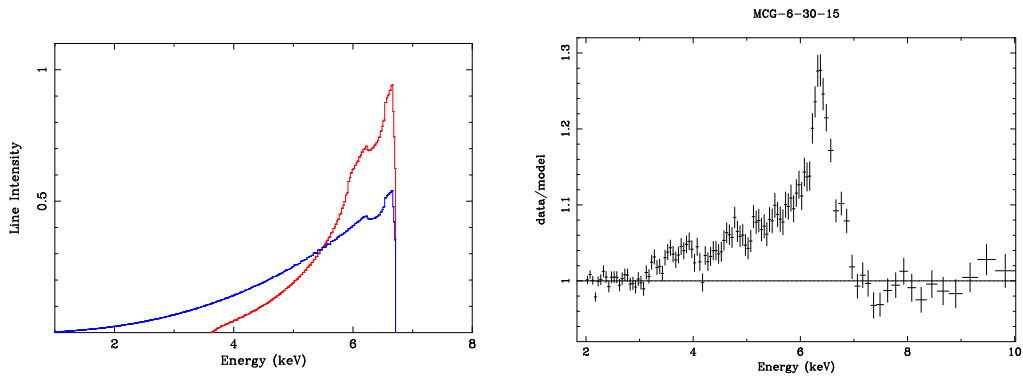} 
\hfill
\includegraphics[height=0.48\textwidth,angle=-90]{mcg6_linebw.ps}
\caption{Left: The line profile dependence on the inner disc
      radius is shown for the two extremal cases of a Schwarzschild BH
      (red, with inner disc radius at $6~r_g$) and of a Maximal Kerr
      BH (blue, with inner disc radius at $\simeq 1.24~r_g$). Right: The
  broad iron line in MCG--6-30-15 from the XMM observation in 2001 (Fabian
  {\etal}\ 2002a) is shown as a ratio to the continuum model.}
\end{figure}

\begin{figure}[tbh]
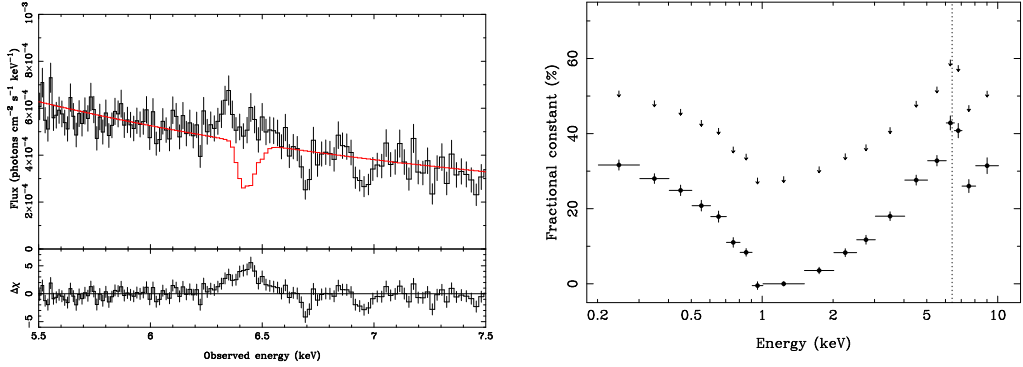

\includegraphics[height=0.48\textwidth,angle=-90]{young7.ps} 
\hfill
\includegraphics[height=0.47\textwidth,angle=-90]{mcgfrac.ps}
  \caption{Left: The Chandra HEG spectrum overlaid with an ionized absorber
    model for the red wing (Young \etal\ 2005). Note that the
    absorption between 6.4 and 6.5~keV predicted by the absorber model
    is not seen. Right: The fractional spectrum of the constant component of
    MCG--6-30-15, constructed from the intercept in flux--flux plots
    (Vaughan \& Fabian 2004). This component strongly resembles
    reflection.}
\end{figure}

\begin{figure}[tbh]
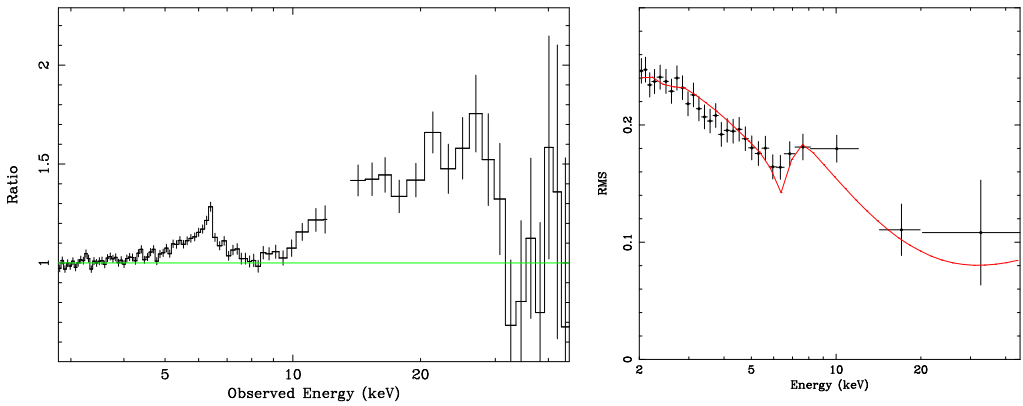

\includegraphics[width=0.39\textwidth,angle=-90]{RatioMCG6.ps}
\hfill
\includegraphics[width=0.38\textwidth,angle=-90]{fvar.ps}
\caption{Left: Broad iron line and Compton hump in MCG--6-30-15 from
  Suzaku. Right: Fractional r.m.s. variability as a function of
  energy compared with model (solid line) in which the power-law is
  assumed to change in normalization whilst the reflection component
  is fixed.}
\end{figure}

All three main parts of the reflection spectrum have now been seen
from AGN and Galactic Black Holes (GBH). The broad iron line and
reflection hump are clearly seen in the Seyfert galaxy MCG--6-30-15
and in the GBH J1650-400. More recently it has been realised
that the soft excess in many AGN can be well explained by smeared
reflection (Crummy \etal\ 2006). It had been noted by Czerny \etal\
(2003) and by Gierlinski \& Done (2004) that the soft excess posed a
major puzzle if thermal since the required temperature was always
about 150~eV, irrespective of black hole mass, luminosity etc.
Explaining it as a feature due to smeared atomic lines resolves this
puzzle. An alternative interpretation involves smeared absorption
lines (Gierlinski \& Done 2004).

The extent of the blurring of the reflection spectrum is determined by
the innermost radius of the disk (Fig.~3). Assuming that this is the
radius of marginal stability enables the spin parameter $a$ of the
hole to be measured. Objects with a very broad iron line like
MCG--6-30-15 are inferred to have high spin $a>0.95$ (Dabrowski \etal\
1997; Brennemann \etal\ 2006). Some (Krolik \& Hawley 2002) have argued
that magnetic fields in the disk can blur the separation between
innermost edge of the disk and the inner plunge region so that the
above assumption is invalid. This probably makes little difference for
the iron line however since the low ionization parameter of most
observed reflection reflection requires that the disc matter is very
dense. The density of matter in the plunge region drops very rapidly
to a low values (Reynolds \& Begelman 1997) and only very strong
magnetic fields, much larger than are inferred in disks, can stop this
steep decline in density. Any reflection from the plunge region will
be very highly ionized and so produce little iron emission.

Broad iron lines and reflection components are seen in both AGN (e.g.
Tanaka \etal\ 1995; Nandra \etal\ 1997) and GBH (Martocchia \& Matt
2002; Miller \etal\ 2002abc, 2003ab, 2004ab). A recent exciting
development are the reports that broad iron lines are present in 50 per
cent of all XMM AGN observations where the data are of high quality
(more than 150,000 counts; Guanazzi \etal\ 2006; Nandra \etal\ 2006). 
They are not found in all objects or in all accretion states. There are
many possible reasons for this, including overionization of the surface,
low iron abundance, and beaming of the primary power-law away from the
disk (e.g. if the power-law originates from the mildly-relativistic base
of the jet).

In many cases there is a narrow iron line component due to reflection
from distant matter. Absorption due to intervening gas, warm absorbers
and outflows from the AGN, as well as the interstellar medium in both
our Milky Way galaxy and the host galaxy must be accounted for.
Moreover, if most of the emission emerges from within a few
gravitational radii and the abundance is not high, then the extreme
blurring can render the blurred reflection undetectable (Fabian \&
Miniutti 2005).

\begin{figure}[tbh]
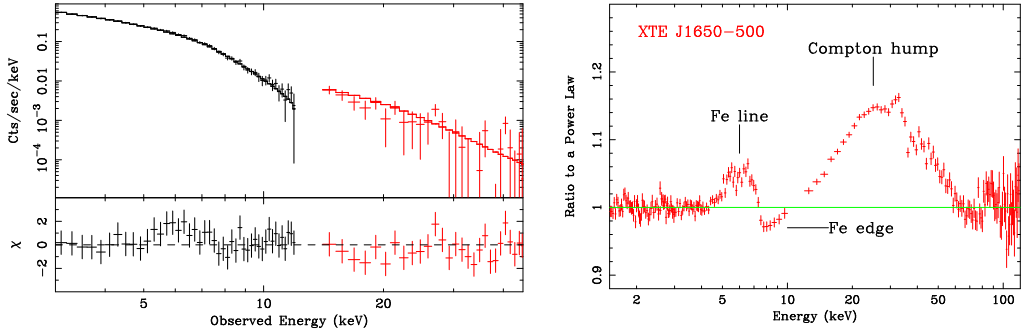

\includegraphics[height=0.51\textwidth,angle=-90]{Diff.ps}
\hfill
\includegraphics[height=0.45\textwidth,angle=-90]{gio1650.ps}
\caption{The broadband BeppoSAX spectrum of XTE J1650--500 (pltotted as a
  ratio to the continuum). The signatures of relativistically-blurred
  reflection are clearly seen (Miniutti \etal\ 2004).}
\end{figure}

\begin{figure}[tbh]
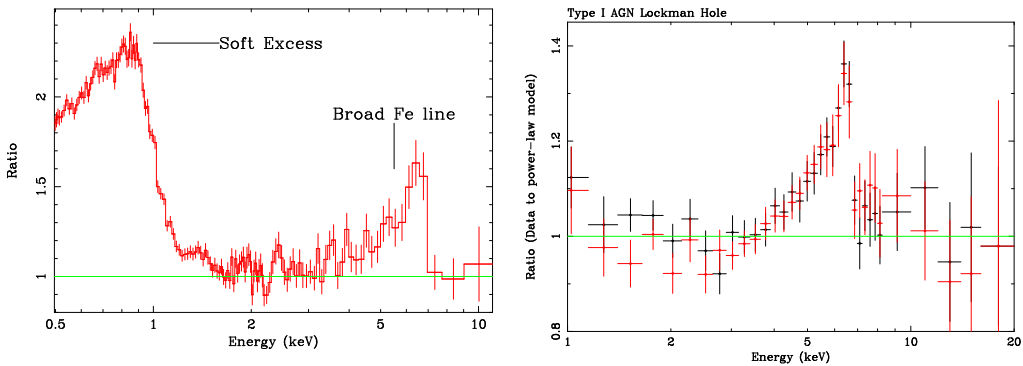
 
\includegraphics[height=0.48\textwidth,angle=-90]{1h07line.ps}
\hfill
\includegraphics[height=0.49\textwidth,angle=-90]{alina_new.ps}
\caption{Left: Ratio of the spectrum of the NLS1 galaxy 1H0707 to a
    power-law. Spectral fits with either a very broad iron line or a
    partial covering with a steep edge are equally good for this
    object (Fabian \etal\ 2004; Boller \etal\ 2002). Right: Ratio plot of the
    mean unfolded spectrum for type--1 AGN in the Lockman Hole with
    respect to a power law (Streblyanska \etal\ 2005).}
\end{figure}

In order to distinguish between the various spectral components, both
emission and absorption, we can use higher spectral resolution,
broader bandwidth and variability. An example of the use of higher
spectral resolution is the work of Young \etal\ (2005) with the Chandra
high energy gratings (Fig.~4). Observations of MCG--6-30-15 fail to show
absorption lines or feature associated with iron of intermediate
ionization. Such gas could cause some curvature of the
apparent continuum mimicking a very broad line. A broader bandwidth is
very useful in determining the slope of the underlying continuum. This
has been shown using BeppoSAX (e.g. Guainazzi \etal\ 1999) and now in
several sources with Suzaku (Fig. 5, Miniutti \etal\ 2006; Reeves \etal\
2006).

\section{Variability}

In the best objects where a very broad line is seen (e.g. MCG--6-30-15 --
Fabian \etal\ 2002; NGC4051 -- Ponti \etal\ 2006) the reflection appears to
change little despite large variations in the continuum. The spectral
variability can be decomposed into a highly variable power-law and a
quasi-constant reflection component.  This behaviour is also borne out
by a difference (high-low) spectrum which is power-law in
shape and the reflection-like shape of the spectrum of the intercept in
flux-flux plots (Figs. 6--7).

This behaviour was initially puzzling, until the effects of
gravitational light bending were included (Fabian \& Vaughan 2003;
Miniutti \etal\ 2004; Miniutti \& Fabian 2005). Recall that the extreme
blurring in these objects means that much of the reflection occurs
with a few $r_{\rm g}$ of the horizon of the black hole. The enormous
spacetime curvature there means that changes in the position of the
primary power-law continuum have a large effect on the flux seen by an
outside observer (Martocchia \& Matt 1996; Martocchia \etal\ 2002).
What this means is that an intrinsically constant continuum source can
appear to vary by large amounts just be moving about in this region of
extreme gravity. The reflection component, which comes from the
spatially fixed accretion disc, appears relatively constant in flux in
this region.  Consequently, the observed behaviour of these objects
may just be a consequence of strong gravity.

Some of the Narrow-Line Seyfert 1 galaxies such as 1H0707, IRAS13224
and 1H0439 appear to share this behaviour (Fabian \etal\ 2002b, 2004,
2005) and can be interpreted in terms of extreme light bending.  Some
of these objects can show sharp drops around 7~keV that may be
alternatively interpreted as due to absorption from something only
partially covering the source (Boller et al 2002). (If the covering was total then no
strong soft emission would be seen, contrary to observation.) The GBH
XTE\,J1650-500 behaved in a manner similar to that expected from the
light bending model (Rossi \etal\ 2005).

\section{Polarization}

The X-ray emission from accreting black holes is expected to be
polarized (Rees 1975), with general relativistic effects 
influencing the degree and angle of polarization of emission from the
innermost regions (Stark \& Connors 1977; Dov\v{c}iak \etal\ 2004,
Fig.~8).
Hopefully, a mission in the near future will carry a sensitive
polarimeter so that this powerful information channel can be opened up.

\begin{figure}[tbh]
\includegraphics[width=0.49\textwidth,angle=0]{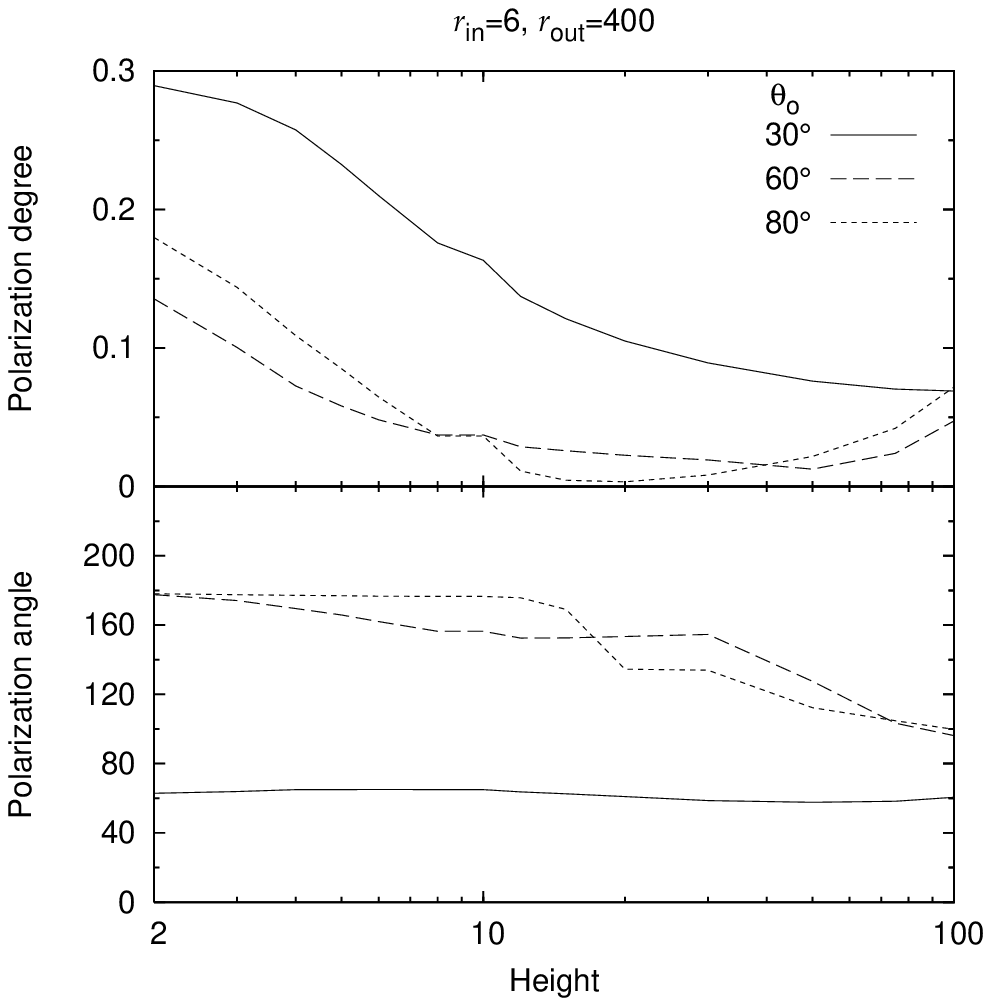}
\hfill
\includegraphics[width=0.49\textwidth,angle=0]{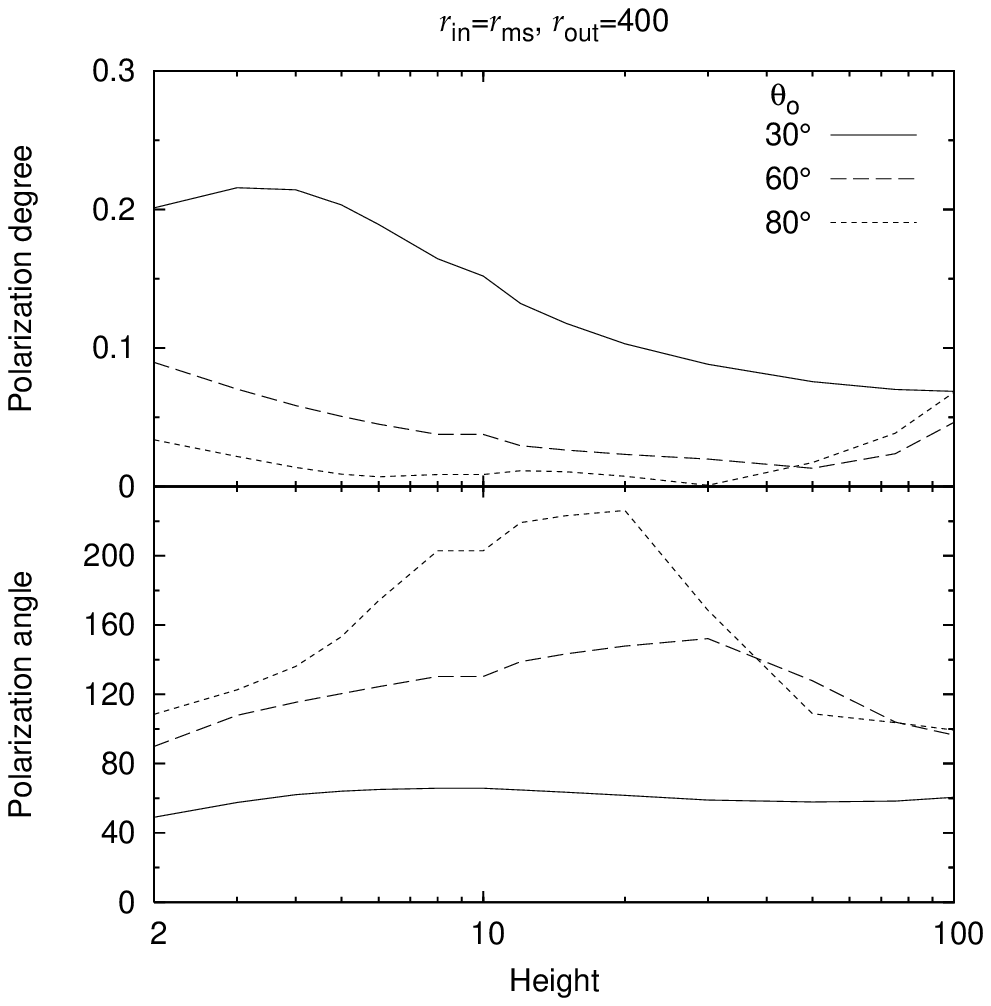}
\caption {Polarization properties in the 9--12~keV band expected from
  a disc around a rapidly spinning black hole as a function of the
  height of the source above the disc in gravitational radii (Dov\v{c}iak,
  Karas \& Matt 2004). The different lines are for different
  inclination angles. The inner radius of the disc is $6r_{\rm g}$ on
  the left and $1.2r_{\rm g}$ on the right.}
\end{figure}

\section{Discussion}

Clear examples of relativistically-broadened iron lines are seen in
some AGN and GBH in some states. Such objects must have dense inner
accretion disks in order that the gas is not overionized. Detection of
a line is helped greatly if the iron abundance is super-Solar and if
there is little extra absorption due to very strong warm absorbers or
winds. Where broad lines are seen and can be modelled satisfactorally
then the spin of the central black hole can be reliably determined.

The study of absorption and emission variability of iron-K lines is in
its infancy, with some interesting and tantalising results produced so
far. The inner regions of accretion flows are bound to be structured
and so give rise to variations. Some may be due to motion or
transience in the corona or primary power-law source while others may
reflect structure, e.g. spiral waves, on the disk itself, intercepting
primary radiation from much smaller radii. 

The number of broad lines detected is increasing and will continue to
expand with the improved broad band coverage from Suzaku.  Hints that
broad lines are common in fainter objects such as in the Lockman Hole
(Streblyanska \etal\ 2005) and Chandra Deep Fields (Brusa \etal\
2005) could indicate that the conditions necessary for strong line
production, perhaps high metallicity, are common in typical AGN at
redshifts 0.5--1.

X-ray astronomers have an excellent tool with which to observe the
innermost regions of accretion disks immediately around spinning black
holes. The effects of redshifts and light bending expected from strong
gravity in this regime are clearly evident. This can and should be
exploited by future X-ray missions. To make significant progress we
need large collecting areas.  The count rate in the broad iron line of
MCG--6-30-15 is about 2~ph~m$^{-2}$~s$^{-1}$, which means that square
metres of collecting area are required around 6~keV in order to look
for reverberation effects. GBH are much brighter but the orbital
periods of matter close to the black hole are much, much smaller so
reverberation is difficult here. Instead, variations with mass
accretion rate and source state are accessible.

\begin{acknowledgements}
I am grateful to Vladim\'{\i}r Karas and Giorgio Matt for proposing and
organising an interesting and successful 
Symposium. I am also grateful to many colleagues for work on broad iron
lines, including Giovanni Miniutti, Jon Miller, Chris Reynolds, Andy
Young, Kazushi Iwasawa, Randy Ross, Simon Vaughan, Luigi Gallo, Thomas
Boller, Jamie Crummy, Josefin Larsson, and to The Royal Society for
continued support. 

This brief review is an update of one I presented
at the ESAC meeting on Variable and Broad Iron Lines.
\end{acknowledgements}

\medskip

\end{document}